\documentclass[12pt,preprintnumbers,nofootinbib,letterpaper]{article}
\pdfoutput=1
\usepackage{amsmath,amssymb,amsfonts,cite,color,epsf,epsfig,epstopdf,graphics,graphicx,mathtools,subcaption,physics,verbatim}
\def\thefootnote{*\arabic{footnote}}
\definecolor{ultramarine}{rgb}{0.07, 0.04, 0.56}
\definecolor{cadmiumgreen}{rgb}{0.0, 0.42, 0.24}
\definecolor{indigo(dye)}{rgb}{0.0, 0.25, 0.42}
\usepackage[linktocpage=true,breaklinks]{hyperref}
\hypersetup{
	colorlinks=true,
	citecolor=ultramarine,
	linkcolor=cadmiumgreen,
	urlcolor=indigo(dye),
}

\usepackage[utf8]{inputenc}

\numberwithin{equation}{section}

\usepackage{array}
\newcolumntype{P}[1]{>{\centering\arraybackslash}p{#1}}

\usepackage[shortlabels]{enumitem}

\usepackage{dcolumn}
\newcolumntype{M}[1]{>{\centering\arraybackslash}m{#1}}
\newcolumntype{N}{@{}m{0pt}@{}}

\usepackage{amsmath}

\newcommand{\Mpl}{M_{\rm Pl}}
\newcommand{\cs}{c_s}
\newcommand{\cv}{c_v}
\newcommand{\ct}{c_t}

\newcommand{\D}{{\rm d}}

\newcommand{\be}{\begin{equation}}  
\newcommand{\ee}{\end{equation}}

\usepackage{tikz}
\begin{document}


\begin{center}

\def\thefootnote{\fnsymbol{footnote}}

\vspace*{1.5cm}
{\Large {\bf Spin-2 dark matter from inflation}}
\\[1cm]

{Mohammad Ali Gorji\footnote{gorji@icc.ub.edu} }
\\[.7cm]

{\small \textit{Departament de F\'{i}sica Qu\`{a}ntica i Astrof\'{i}sica, Institut de Ci\`{e}ncies del Cosmos, Universitat de Barcelona, Mart\'{i} i Franqu\`{e}s 1, 08028 Barcelona, Spain
}}

\end{center}

\vspace{1cm}

\hrule \vspace{0.5cm}

\begin{abstract}
The seed of dark matter can be generated from light spectator fields during inflation through a similar mechanism that the seed of observed large scale structures are produced from the inflaton field. The accumulated energy density of the corresponding excited modes, which is subdominant during inflation, dominates energy density of the universe later around the time of matter and radiation equality and plays the role of dark matter. For spin-2 spectator fields, Higuchi bound may seem to prevent excitation of such light modes since deviation of the inflationary background from the exact de Sitter spacetime is very small. However, sizable interactions with the inflaton field breaks (part of) isometries of the de Sitter space in the inflationary background and relaxes the Higuchi bound. Looking for this possibility in the context of effective field theory of inflation, we suggest a dark matter model consisting of spin-2 particles that produce during inflation.
\end{abstract}
\vspace{0.5cm} 

\hrule
\def\thefootnote{\arabic{footnote}}
\setcounter{footnote}{0}

\thispagestyle{empty}


\newpage

\section{Introduction}\label{introduction}
With yet unknown origin, there are many beyond standard model candidates for the dark matter (DM) particles (see  \cite{Bertone:2004pz,Feng:2010gw} and references therein). Based on our current understanding, DM particles should be produced sometime before the time of matter and radiation equality, i.e., during the radiation domination, reheating or even inflation. However, most of the models deal with production of DM particles during radiation domination and not inflation. Indeed, even if some particles are produced during inflation, the corresponding energy density would quickly dilute due to the exponential expansion of the inflationary universe. Nevertheless, there has been some attempts to construct models in which DM particles are produced during inflation \cite{Polarski:1994rz,Chung:2001cb,Graham:2015rva,Kolb:2017jvz,Bastero-Gil:2018uel,Nakayama:2019rhg,Nakayama:2020rka,Nakai:2020cfw,Salehian:2020asa,Firouzjahi:2020whk,Firouzjahi:2021lov,Bastero-Gil:2021wsf,Sato:2022jya,Redi:2022zkt,Bastero-Gil:2022fme,Kolb:2022eyn,Nakai:2022dni,Kitajima:2023fun}. In the case of vector fields, direct coupling with inflaton is necessary to produce spin-1 DM particles \cite{Graham:2015rva,Bastero-Gil:2018uel,Nakayama:2019rhg,Nakayama:2020rka,Nakai:2020cfw,Salehian:2020asa,Firouzjahi:2020whk,Bastero-Gil:2021wsf,Sato:2022jya,Redi:2022zkt,Bastero-Gil:2022fme,Nakai:2022dni,Kitajima:2023fun} which is usually called vector DM. On the other hand, scalar fields can provide spin-0 DM particles which are produced through the gravitational interaction at superhorizon scales in a similar way that curvature perturbations and metric tensor perturbations (primordial gravitational waves) produce during inflation \cite{Polarski:1994rz,Graham:2015rva,Firouzjahi:2020whk,Firouzjahi:2021lov,Kolb:2022eyn}. This is an interesting idea that the seed of DM particles are produced during inflation through an almost similar mechanism that the seed of the observed large scale structures in the universe are produced. However, the situation is more subtle for spin-2 fields with which we are interested in this paper.

It is well known that spin-0 (scalar) light spectator fields, which are subdominant, can be produced at the superhorizon scales during inflation. However, it is not easy to produce fields with higher spin during inflation due to the high symmetry group of the de Sitter (dS) spacetime. At superhorizon scales, perturbations of a field with spin ${\rm s}$ and mass $m$ in dS spacetime scales as $(-\tau)^{\Delta_{\pm}}$ where $\tau$ is the conformal time and $\Delta_{\pm}$ is the scaling dimension which is given by \cite{Arkani-Hamed:2015bza,Bordin:2016ruc,Kehagias:2017cym}
\begin{align}
\Delta_{\pm} = \frac{3}{2} \pm \sqrt{\left({\rm s}-\frac{1}{2}\right)^2-\frac{m^2}{H^2}} \,,
\end{align}
where $H$ is the Hubble constant characterizing constant curvature of the dS spacetime. At superhorizon scales $\tau\to0^{-}$, $\Delta_-<1$ is needed to prevent decay of the perturbations of fields with higher spin. In particular, $m^2<0$ and $m^2<2H^2$ are needed for spin-1 ${\rm s}=1$ and spin-2 ${\rm s}=2$ modes respectively. More importantly, the spin-0 mode ${\rm s}=0$ of a spinning field immediately becomes ghost for $\Delta_-<1$. The so-called Higuchi bound \cite{Higuchi:1986py}
\begin{align}\label{Higuchi-bound}
m^2 > {\rm s} ({\rm s}-1) H^2 \,,
\end{align}
prevents $\Delta_-<1$ to guaranty the absence of such a ghost for the longitudinal mode of a higher spin field. For example, for a spin-2 field, $m^2<2H^2$ is needed to prevent decay of the superhorizon perturbations while $m^2>2H^2$ is needed by Higuchi bound \eqref{Higuchi-bound} to have a healthy longitudinal mode. In this regards, at the level of perturbations, fields with higher spin decay at least as $a^{-1}$ at superhorizon scales and it is not possible to produce long-lived particles with higher spin during inflation.

On the other hand, it is also known that if a spin-1 (vector) field has large enough coupling with inflaton, it is possible to produce long-lived vector perturbations \cite{Ratra:1991bn,Watanabe:2009ct}. This is simply because of the fact that the large coupling breaks part of the dS symmetries. In the case of vector field, the conformal invariance breaks due to the non-minimal kinetic coupling. Similarly, one may relax the Higuchi bound \eqref{Higuchi-bound} for spin-2 and higher spin fields by considering large enough couplings with the inflaton field. This possibility is recently studied by implementing different approaches \cite{Kehagias:2017cym,Baumann:2017jvh,Goon:2018fyu,Bordin:2018pca}. Having long-lived higher spin fields during inflation, their observable effects on the CMB spectrum is studied when they are considered to be mediators (as internal legs in Feynman diagrams) for the correlation functions of the curvature perturbations and primordial tensor perturbations \cite{Baumann:2017jvh,Goon:2018fyu,Bordin:2018pca,Dimastrogiovanni:2018gkl,Bordin:2019tyb,Iacconi:2019vgc,Iacconi:2020yxn}. Here, however, we are interested in their direct effects (as external legs in Feynman diagrams). We will show that a spin-2 field during inflation can be considered as a DM candidate. The idea of spin-2 and also higher spin DM models is already studied in the literature. The models either deal with the heavy (compared with the Hubble parameter during inflation) DM particles produced during inflation \cite{Alexander:2020gmv,Falkowski:2020fsu,Kolb:2023dzp} or 
deal with production of DM after inflation \cite{Maeda:2013bha,Aoki:2014cla,Aoki:2016zgp,Babichev:2016bxi,Aoki:2017cnz,Aoki:2017ffl,Marzola:2017lbt,Chu:2017msm,GonzalezAlbornoz:2017gbh,Criado:2020jkp,Falkowski:2020fsu,Jain:2021pnk,Manita:2022tkl,Wu:2023dnp}. One simple reason is the restriction from the Higuchi bound \eqref{Higuchi-bound}. Considering a large enough coupling with the inflaton and relaxing this strict bound, we look for production of spin-2 DM particles during inflation.

The rest of the paper is organized as follows. In Sec. \ref{app-EFT}, we construct the effective field theory (EFT) of a light spin-2 field during inflation. Focusing on the subset of the general EFT, we present our model in Sec. \ref{sec-model}. In Sec. \ref{sec-spin2-DM}, we investigate the necessary conditions that spin-2 particles should satisfy to be a viable DM candidate. We present some constraints on the model in Sec. \ref{sec-constraints} and we estimate the relic density of DM in Sec. \ref{sec-DM-relic}. Sec. \ref{summary} is devoted to the summary of the results.

\section{EFT of a light spin-2 field during inflation}\label{app-EFT}

In this section, we present a brief review of the EFT of a light spin-2 field during inflation. The readers who are not interested in the details of EFT construction may directly move to the next section.

The background configuration of the single field inflation is defined by spatially flat FLRW spacetime and time-dependent inflaton field
\begin{align}\label{BG}
&{\bar g}_{\mu\nu} dx^\mu dx^\nu = - N(t)^2 dt^2 + a(t)^2 \delta_{ij} dx^i dx^j \,,
&{\bar \phi}(t) \,,
\end{align}
where $N(t)$ is the lapse function, $a(t)$ is the scale factor, $\phi$ is the inflaton field and a bar denotes the corresponding background value.

Time dependency of the inflaton field in \eqref{BG} spontaneously breaks time diffeomorphisms leaving time-dependent spatial diffeomorphisms as the residual symmetries. The EFT of single field inflationary scenarios is based on the systematic construction of all possible operators which respect this symmetry breaking pattern in unitary gauge where $\phi=t$ \cite{Cheung:2007st}. Technically, inflaton field $\phi$ defines time slices which can be characterized by the following timelike vector
\begin{align}\label{n}
&n_{\mu} = - \frac{\partial_\mu\phi}{\sqrt{-g^{\alpha\beta}\partial_\alpha\phi\partial_\beta\phi}} \,;
&n_\mu n^\mu = -1 \,.
\end{align}
One then looks for all operators which live on the three-dimensional timelike hypersurfaces defined by the above normal vector with induced metric
\begin{align}\label{h}
h_{\mu\nu} = g_{\mu\nu} + n_\mu n_\nu \,,
\end{align}
that satisfies $h^\mu{}_\alpha n^\alpha=0$. 

Using the same technique, one can extend the EFT of single field inflation to include light spin-0 (scalar) degrees of freedom other than inflaton \cite{Senatore:2010wk}. As we explained in Sec.~\ref{introduction}, including light fields with higher spin is more subtle. One reason is that the mass of the higher spin field in dS background is restricted to the Higuchi bound \eqref{Higuchi-bound}. However, this bound can be relaxed if higher spin fields have sizable interactions with the inflaton field. More precisely, the Higuchi bound arises due to the symmetries of the dS space and by having sizable interactions with inflaton we may deviate enough from the exact symmetries of dS spacetime. In this regard, one can construct an EFT for the fields with higher spin. 

The five degrees of freedom of spin-2 field are encoded in symmetric traceless spatial three-tensor $\Sigma^{ij}$. Alternatively, one can work with symmetric traceless spatial four-tensor $\Sigma^{\mu\nu}$ which satisfies $n_\mu\Sigma^{\mu\nu}=0$ and has the following components \cite{Bordin:2018pca}
\begin{align}\label{Sigma-components}
&\Sigma^{00} = \frac{\partial_i\pi\partial_j\pi}{(1+\dot{\pi})^2} \Sigma^{ij}\,, 
&\Sigma^{0i} = -\frac{\partial_j\pi}{1+\dot{\pi}} \Sigma^{ij} \,,
\end{align} 
where $\phi = t + \pi(t,{\bf x})$ is used. Note that $\pi$ is the perturbation of the inflaton and $\Sigma^{\mu\nu}$ has five independent components as desired for a spin-2 field. In unitary gauge $\pi=0$, $\Sigma^{00}=0=\Sigma^{0i}$ and $\Sigma^{\mu\nu}$ has only pure spatial nonzero components $\Sigma^{ij}$.

The free EFT action for light spin-2 field $\Sigma_{\mu\nu}$ is given by\footnote{Alternatively, one could consider $B\,h^{\mu\nu}\nabla_\mu \Sigma^{\alpha\beta} \nabla_\nu \Sigma_{\alpha\beta}$ instead of $B\, g^{\mu\nu} \nabla_\mu \Sigma^{\alpha\beta} \nabla_\nu \Sigma_{\alpha\beta}$ in \eqref{action-free-general} so that the $B$ term would not provide any time derivative of $\Sigma_{\alpha\beta}$.}
\begin{align}\label{action-free-general}
S_{\Sigma} &= \frac{1}{4} \int d^4x \sqrt{-g} \Big[ 
A\, n^\mu n^\nu \nabla_\mu \Sigma^{\alpha\beta} \nabla_\nu \Sigma_{\alpha\beta}
- B \, \nabla_\mu \Sigma^{\alpha\beta} \nabla^\mu \Sigma_{\alpha\beta}
- C\, \nabla_\mu \Sigma^{\mu\alpha} \nabla^\nu \Sigma_{\nu\alpha}
- D\, \Sigma^{\alpha\beta} \Sigma_{\alpha\beta}
\Big] \,,
\end{align}
where $A, B, C, D$ are unknown functions of time. The time dependency of these functions arise due to the direct interactions between inflaton $\phi$, which is only a function of time in unitary gauge, and spin-2 field $\Sigma^{\mu\nu}$. Thus, \eqref{action-free-general} is not really a free action in this sense. It is called free action in the sense that it includes leading operators in derivative expansion. The next leading order operators determine the interaction action \cite{Bordin:2018pca}
\begin{align}\label{action-int-general}
S_{\rm int} &= \int d^4x \sqrt{-g} \Big[ 
\kappa\Mpl \, \delta{K}_{\mu\nu} \Sigma^{\mu\nu} 
+ \tilde{\kappa} \Mpl \, \delta g^{00} \delta{K}_{\mu\nu} \Sigma^{\mu\nu}
- \mu\, \Sigma^{\mu\nu} \Sigma_{\mu}{}^\alpha \Sigma_{\alpha\nu}
\Big] \,,
\end{align}
where  $\kappa, \tilde{\kappa}, \mu$ are unknown functions of time and $\delta{g}^{00}=g^{00}-\bar{g}^{00}$ and $\delta{K}_{\mu\nu} = K_{\mu\nu} - a^2 H h_{\mu\nu}$ characterize perturbations encoded in $g^{00}$ and extrinsic curvature. 

The full action of the theory includes action of inflaton $S_\phi$, which we assume that is minimally coupled to the Einstein-Hilbert action, \eqref{action-free-general} and \eqref{action-int-general}. Variation of the full action with respect to the metric gives the Einstein equations
\begin{align}\label{EEs}
&\Mpl^2 G_{\mu\nu} = T_{\mu\nu} \,;
&T_{\mu\nu} = \frac{-2}{\sqrt{-g}}\frac{\delta}{\delta{g}^{\mu\nu}}\left(S_{\phi}+ S_{\Sigma}+S_{\rm int}\right) \,,
\end{align}
where $T_{\mu\nu}$ is the total energy-momentum tensor. The energy-momentum tensor takes a complicated form and we will not write here its explicit form.

Variation with respect to $\Sigma^{\mu\nu}$ gives equation of motion for the spin-2 field
\begin{align}
\nabla_\alpha\left( A\, n^\alpha n^\beta \nabla_\beta \Sigma_{\mu\nu} 
- B\, \nabla^\alpha \Sigma_{\mu\nu} - C\, \delta^\alpha{}_\mu \nabla^\beta \Sigma_{\beta\nu} \right) 
+ D\, \Sigma_{\mu\nu} + 6 \mu\, \Sigma_{\mu}{}^\alpha \Sigma_{\alpha\nu}
=  2 \Mpl \left( \kappa+\tilde{\kappa}\delta{g}^{00} \right) \delta{K}_{\mu\nu} \,.
\end{align}

The free action \eqref{action-free-general} takes the following form
\begin{align}\label{action-free-Sigma-p}
S_{\Sigma} &= 
\frac{1}{4} \int \D^3x\, \D{t}\, { N} a^3 \left[ 
(A+B) \left( \dot{\Sigma}_{ij} \right)^2 
- \frac{B}{a^2} \left( \partial_i{\Sigma}_{jk} \right)^2
- \frac{C}{a^2} \left( \partial_i{\Sigma}^{ij} \right)^2
- \left(D-2B H^2\right) \left( {\Sigma}_{ij} \right)^2
\right] \,,
\end{align}
where a dot is defined as $\dot{}\equiv{d}/{{ N}dt}$ such that it reduces to derivative with respect to cosmic time $t$ and conformal time $\tau=\int\D{t}/a(t)$ for $N=1$ and $N=a$ respectively. 

The energy density of the spin-2 field is given by $\rho_\Sigma=-T^{0}_{{\Sigma}\,0}$, where $T_{{\Sigma}\,\mu\nu} = \frac{-2}{\sqrt{-g}}\frac{\delta{S}_{\Sigma}}{\delta{g}^{\mu\nu}}$ is the energy-momentum tensor of the spin-2 field. With straightforward calculations we find
\begin{align}\label{energy-density-sigma}
\rho_\Sigma &= 
\frac{1}{2} \left[ 
(A+B)\left( \dot{\Sigma}_{ij} \right)^2 
+ \frac{B}{a^2}  \ct^2 \left( \partial_i{\Sigma}_{jk} \right)^2
+ \frac{C}{a^2 } \left( \partial_i{\Sigma}^{ij} \right)^2
+ \left( D + 2B H^2 \right) \left( {\Sigma}_{ij} \right)^2
\right] - \frac{2C}{a^2} \partial_iJ^i \,,
\end{align}
where $J^i = \Sigma^{ij} \partial_k \Sigma_j{}^{k}$.

\section{The model}\label{sec-model} 

We focus on the following subset of the general EFT that is constructed in the previous section
\begin{align}\label{parametrization}
&A = 1 - \ct^2\,, 
&&B = \ct^2\,, 
&&C = \frac{3}{2}\left( \cs^2 - \ct^2 \right)\,, &&D = m^2 + 2 \ct^2 H^2\,.
\end{align}
The free action \eqref{action-free-Sigma-p} then simplifies to\footnote{Note that $c_t$ is the sound speed for the tensor perturbations of the spin-2 field which is different than the speed of gravitational waves which are characterized by the metric tensor perturbations.} 
\begin{align}\label{action}
S_\Sigma &= 
\frac{1}{4} \int \D^3x\, \D{t}\, { N} a^3 \left[ 
\left( \dot{\Sigma}_{ij} \right)^2 
- \frac{1}{a^2}  \ct^2 \left( \partial_i{\Sigma}_{jk} \right)^2
- \frac{3}{2a^2} \left(\cs^2-\ct^2\right) \left( \partial_i{\Sigma}^{ij} \right)^2
- m^2 \left( {\Sigma}_{ij} \right)^2
\right] \,.
\end{align}
The functions $\cs, \ct, m$ are functions of time in general. However, for the sake of simplicity, we assume that their time dependency is negligible such that we can treat them as constants. Furthermore, the parametrization \eqref{parametrization} corresponds to the canonical normalization of the spin-2 field and, as we will show explicitly by decomposing the spin-2 field in this section, it is chosen such that parameters $\cs$, $\cv$, $\ct$, where
\begin{align}\label{cv}
\cv^2 \equiv \frac{3}{4} \cs^2 + \frac{1}{4} \ct^2 \,,
\end{align} 
are sound speeds for the scalar, vector, tensor perturbations of the spin-2 field. 

Note that there are four independent functions $A, B, C, D$ in general while there are only three independent functions $\cs, \ct, m$ in our model \eqref{action} since we have assumed parameterization \eqref{parametrization}. In other words, $\cv$ could be a free parameter in general and relation \eqref{cv} holds due to the parameterization \eqref{parametrization}. Indeed, we are neither forced to choose this parameterization nor to assume $\cs, \ct, m$ are constants and we could keep general time-dependent parameters $A, B, C, D$. However, this simple parametrization with constant $\cs, \ct, m$ is enough for our purpose in this paper.

Moreover, we need energy density of the spin-2 field which is obtained in Eq. \eqref{energy-density-sigma}. For the parametrization \eqref{parametrization}, we find
\begin{align}\label{rho-sigma}
\rho_\Sigma &= 
\frac{1}{2} \left[ 
\left( \dot{\Sigma}_{ij} \right)^2 
+ \frac{1}{a^2}  \ct^2 \left( \partial_i{\Sigma}_{jk} \right)^2
+ \frac{3}{2a^2 }\left(\cs^2-\ct^2\right) \left( \partial_i{\Sigma}^{ij} \right)^2
+ \left( m^2 + 4 \ct^2 H^2 \right) \left( {\Sigma}_{ij} \right)^2
\right] \,,
\end{align}
where we have ignored the total divergence term shown in Eq. \eqref{energy-density-sigma} since it does not contribute to the divergence-free homogeneous background provided by average of the energy density.

Action \eqref{action} determines dynamics of the spin-2 field in our setup. After solving the dynamics, we can substitute the solution in \eqref{rho-sigma} to find energy density. Before doing so, we decompose the spin-2 field to its five degrees of freedom in the next two subsections. This makes the roles of different degrees of freedom more clear.

\subsection{SVT decomposition}

The massive spin-2 field has five degrees of freedom. In our EFT setup with time-dependent spatial diffeomorphism as residual symmetry, the five degrees of freedom are characterized by one longitudinal scalar mode, two transverse vector modes and two traceless and transverse tensor modes. Thus, we consider the following scalar-vector-tensor (SVT) decomposition
\begin{align}\label{sigma-decomposition}
\Sigma_{ij} = \partial_i\partial_j s - \frac{1}{3} \delta_{ij}\partial^2 s + 2 \partial_{(i}v_{j)} + t_{ij} \,,
\end{align}
so that $s$, $v_i$ and $t_{ij}$ characterize scalar, vector and tensor modes respectively. The vector modes are divergence-free $\partial_iv^i=0$ and the tensor modes are traceless and transverse $\partial_it^{ij}=0=\delta^{ij}t_{ij}$. Note that $\delta^{ij}\Sigma_{ij}=0$ as desired.

Substituting \eqref{sigma-decomposition} in action \eqref{action} and performing some integration by parts, we find
\begin{align}\label{action-total}
S_\Sigma=S_{\rm s}[s]+S_{\rm v}[v]+S_{\rm t}[t] \,,
\end{align}
where the quadratic actions for the scalar, vector, and tensor modes are given by
\begin{align}\label{action-free-S}
S_{\rm s}[s] & = 
\frac{1}{6} \int \D^3x\, \D\tau\, a^2 \left[ 
\left( \partial^2{s}' \right)^2 
- \cs^2 \left( \partial_i\partial^2s \right)^2
- m^2 a^2 \left(\partial^2s\right)^2
\right] \,,
\\ \label{action-free-V}
S_{\rm v}[v_i] & = 
\frac{1}{2} \int \D^3x\, \D\tau\, a^2 \left[ 
\left( \partial_i{v}'_j\right)^2
- \cv^2 \left(\partial_i\partial_j{v}_k \right)^2
- m^2 a^2 \left( \partial_i{v}_j \right)^2
\right]
\,,
\\ \label{action-free-T}
S_{\rm t}[t_{ij}] & =
\frac{1}{4} \int \D^3x\, \D\tau\, a^2 \left[ 
\left( {t}'_{ij} \right)^2 
- \ct^2 \left( \partial_i{t}_{jk} \right)^2
- m^2 a^2 \left( {t}_{ij} \right)^2
\right] \,,
\end{align}
where a prime denotes derivative with respect to the conformal time with $N=a$. Note that scalar, vector, and tensor modes do not mix at the linear level.

One important comment here is in order. It is well-known that if one starts with a spin-2 field in a dS space in a full covariant way, when there is a spacetime diffeomorphisms, stability of the longitudinal mode leads to the Higuchi bound \eqref{Higuchi-bound} \cite{Higuchi:1986py}. However, as we can clearly see from action \eqref{action-free-S}, the stability of the longitudinal mode $s$ does not imply the Higuchi bound. This is a direct consequence of the fact that we have three independent parameters $\cs, \ct, m$ in our model thanks to the breaking of the full spacetime diffeomorphisms to the spatial diffeomorphisms through the sizable interactions between the spin-2 field and inflaton. Otherwise, isometries of the dS would imply relations between $\cs, \ct, m$ such that they would not be independent parameters and Higuchi bound \eqref{Higuchi-bound} would arise. This is the key point in our setup which makes it possible to excite light spin-2 particles during inflation.

Substituting \eqref{sigma-decomposition} in \eqref{rho-sigma}, we find the following contributions for the energy density from the scalar, vector, and tensor modes
\begin{align}\label{energy-density-tot}
\rho_\Sigma=\rho_{\rm s}[s]+\rho_{\rm v}[v]+\rho_{\rm t}[t] \,,
\end{align}
where
\begin{align}\label{energy-density-S}
\rho_{\rm s}[s] &= \frac{1}{3a^2} \left[ 
\left( \partial^2{s}' \right)^2 
+ \cs^2 \left( \partial_i\partial^2{s} \right)^2
+ \left(m^2 a^2 + 4 \ct^2 \frac{a'^2}{a^2} \right) \left( \partial^2{s} \right)^2
\right]
\,,
\\ \label{energy-density-V}
\rho_{\rm v}[v_i] &=
\frac{1}{a^2} \left[ 
\left( \partial_i{v}'_j \right)^2 
+ \cv^2 \left( \partial_i\partial_j{v}_{k} \right)^2
+ \left(m^2 a^2 + 4 \ct^2 \frac{a'^2}{a^2} \right) \left( \partial_i{v}_{j} \right)^2
\right]
\,,  
\\ \label{energy-density-T}
\rho_{\rm t}[t_{ij}] &= 
\frac{1}{2a^2} \left[ 
\left( {t}'_{ij} \right)^2 
+ \ct^2 \left( \partial_i{t}_{jk} \right)^2
+ \left(m^2 a^2 + 4 \ct^2 \frac{a'^2}{a^2} \right) \left( {t}_{ij} \right)^2
\right] \,.
\end{align}

\subsection{Helicity representation}

Furthermore, we go to Fourier space and rewrite the vector and tensor modes in terms of polarizations as
\begin{align}\label{s-helicity}
s(\tau,{\bf x}) &= 
\int \frac{\D^3k}{(2\pi)^3} e^{i {\bf k}.{\bf x}}\, s_{\bf k}(\tau) \,,
\\
\label{v-helicity}
v_{i}(\tau,{\bf x}) &= \sum_{\lambda=+,\times} 
\int \frac{\D^3k}{(2\pi)^3} e^{i {\bf k}.{\bf x}}
e^{\lambda}_{i}(\hat{\bf k})\, v^\lambda_{\bf k}(\tau) \,,
\\
\label{t-helicity}
t_{ij}(\tau,{\bf x}) &= \sum_{\lambda=+,\times} 
\int \frac{\D^3k}{(2\pi)^3} e^{i {\bf k}.{\bf x}}
e^{\lambda}_{ij}(\hat{\bf k})\, t^\lambda_{\bf k}(\tau) \,,
\end{align}
where $e_{i}^\lambda(\hat{\bf k})$ and $e_{ij}^\lambda(\hat{\bf k})$ are the usual polarization vector and tensors which characterize the helicity-1 and helicity-2 modes of the vector and tensor perturbations respectively (see appendix A of \cite{Salehian:2020dsf} for some necessary properties of polarization vector and tensors). We define
\begin{align}\label{XI}
X_{\bf k}^I &= \left\{ S_{\bf k}, V_{\bf k}^{+}, V_{\bf k}^{\times}, T_{\bf k}^{+}, T_{\bf k}^{\times} \right\}
\equiv {a} \left\{ \frac{-k^2 s_{\bf k}}{\sqrt{3}}, 
k v_{\bf k}^{+}, k v_{\bf k}^{\times},
\frac{t_{\bf k}^{+}}{\sqrt{2}}, \frac{t_{\bf k}^{\times}}{\sqrt{2}} \right\} \,,
\end{align}
and expand the Fourier amplitudes in terms of the annihilation and creation operators as usual $X^{I}_{\bf k}(\tau) = X^{I}_k(\tau) \, \hat{a}_{\bf k}^{I} + X^{\ast{I}}_k(\tau) \, \hat{a}_{\bf k}^{\dagger{I}}$ where  $X^{I}_k(\tau)=\left\{ S_{ k}, V_{ k}^{+}, V_{ k}^{\times}, T_{ k}^{+}, T_{ k}^{\times} \right\}$ are mode functions, $\hat{{a}}_{\bf k}^{I}$ and $\hat{{a}}_{\bf k}^{\dagger{I}}$ are annihilation and creation operators which satisfy the usual commutation relations $[\hat{{a}}_{\bf k}^{I},\hat{{a}}_{\bf q}^{\dagger{J}}] = (2\pi)^3\delta^{IJ} \delta({\bf k}-{\bf q})$. The action \eqref{action-total} then takes the following form
\begin{align}\label{action-conformal}
S_\Sigma &= \frac{1}{2} \sum_{I=1}^5 
\int \frac{\D^3k}{(2\pi)^3}\, \D\tau\, \left[ 
\big| {X'}_{\bf k}^I \big|^2 
- \omega_{I}^2 \big| X_{\bf k}^I \big|^2
\right] \,,
\end{align}
where we have defined frequencies of scalar, vector, and tensor modes as follows 
\begin{align}\label{omega}
\omega_{I}^2 \equiv c_{I}^2 k^2 + m^2 a^2 - \frac{a''}{a} \,;
\qquad c_I\equiv \left\{ \cs, \cv, \cv, \ct, \ct \right\} \,.
\end{align}
The action \eqref{action-conformal} gives the equations of motion for the mode functions
\begin{align}\label{EoM}
{X''}_k^I + \omega_{I}^2 \, X_k^I = 0 \,.
\end{align}

Now, we turn to the energy density for the spin-2 field Eq. \eqref{energy-density-tot}. Going to the Fourier space, we find
\begin{align}\label{rho-sigma-helicity}
\rho_\Sigma = \frac{1}{a^4} \sum_{I=1}^5 
\int \frac{\D^3k}{(2\pi)^3}\, \left[
\bigg| a \left( \frac{{X}_{k}^I}{a}\right)' \bigg|^2
+ \left( 
c_I^2k^2+ m^2 a^2 + 4 \ct^2 \frac{a'^2}{a^2}
\right) \left| X_{k}^I \right|^2
\right] \,.
\end{align}

For the later convenience, it is better to work with the dimensionless fractional energy density
\begin{align}\label{Omega-sigma}
\Omega_\Sigma 
= \frac{\rho_\sigma}{3\Mpl^2H^2} 
= {\cal C} \left(\frac{H}{2\pi\Mpl}\right)^2 \,,
\end{align}
which we have parametrized in terms of a dimensionless time-dependent quantity ${\cal C}$. This parametrization is very convenient for our later purpose.

\section{Spin-2 dark matter}\label{sec-spin2-DM}

\subsection{Production during inflation}
Looking at Eq. \eqref{omega}, we see that the mass term always provides a positive contribution to $\omega_I^2$. Using $a\simeq-1/H_{\rm inf}\tau$, we find that heavy modes $m\gg{H}_{\rm inf}$ cannot be excited. Therefore, efficient particle production happens only for the light modes
\begin{align}\label{mass-light}
m\ll{H}_{\rm inf} \,. 
\end{align}
It is worth mentioning that the above condition violates the Higuchi bound \eqref{Higuchi-bound} which, by neglecting slow-roll suppressed corrections, implies $m>\sqrt{2}H_{\rm inf}$ for ${\rm s}=2$. However, as we have already mentioned, the Higuchi bound does not hold in our setup since (part of) the isometries of dS group breaks in our EFT setup \cite{Bordin:2018pca}. Note that violation of the Higuchi bound is necessary for us to produce light massive spin-2 particles during inflation.

The mass term in \eqref{omega} can be neglected for the light modes \eqref{mass-light} and we find
\begin{align}\label{omega-inflation}
&\omega_I^2 \simeq c_I^2k^2-\frac{2}{\tau^2} \,;
&\mbox{during inflation} \,.
\end{align}
The modes $-c_I k\tau\lesssim1$ satisfy tachyonic condition $\omega_I^2<0$ and will be excited. This is similar to the mechanism of particle production for both curvature perturbations and primordial gravitational waves in the usual inflationary scenarios. The difference is that in our case, the negative term $-2/\tau^2$ showed up thanks to our parameter choice (canonical normalization) for the EFT coefficients in action \eqref{action} which is only possible by assuming a direct coupling with inflaton (for instance, this term is absent for the vector modes due to the conformal symmetry \cite{Watanabe:2009ct}). In the case of curvature perturbations and primordial gravitational waves, a similar term $-2/\tau^2$ appears but with a pure gravitational origin.

The energy density \eqref{rho-sigma-helicity} of spin-2 particles at the end of inflation is
\begin{align}\label{energy-density}
\rho_{\Sigma,{\rm e}} = \frac{1}{a^4} \sum_{I=1}^5 
\int_{\omega_I^2<0} \frac{\D^3k}{(2\pi)^3}\, \left[
\bigg| a \left( \frac{{X}_{k}^I}{a}\right)' \bigg|^2
+ \left( 
c_I^2k^2+ m^2 a^2 + 4 \ct^2 \frac{a'^2}{a^2}
\right) \left| X_{k}^I \right|^2
\right] \Bigg{|}_{\tau=\tau_{\rm e}} \,,
\end{align}
where we have only taken into account contribution from the tachyonic modes as the other modes will contribute to the vacuum fluctuations and should be renormalized away. 

Now, let us explicitly compute \eqref{energy-density}. The positive frequency solution for \eqref{EoM} with Bunch-Davies initial condition are
\begin{align}\label{mode-function}
&X^I_k = \frac{1}{\sqrt{2c_Ik}} \left(1+\frac{i}{x^I}\right) e^{ix^I} \,;
&x^I\equiv-c_I k \tau \,,
\end{align}
where we have considered light modes \eqref{omega-inflation} and we have also treated $c_I$ as constants.

\begin{figure}[htbp!]
	\centering
	\includegraphics[width=.85 \columnwidth]{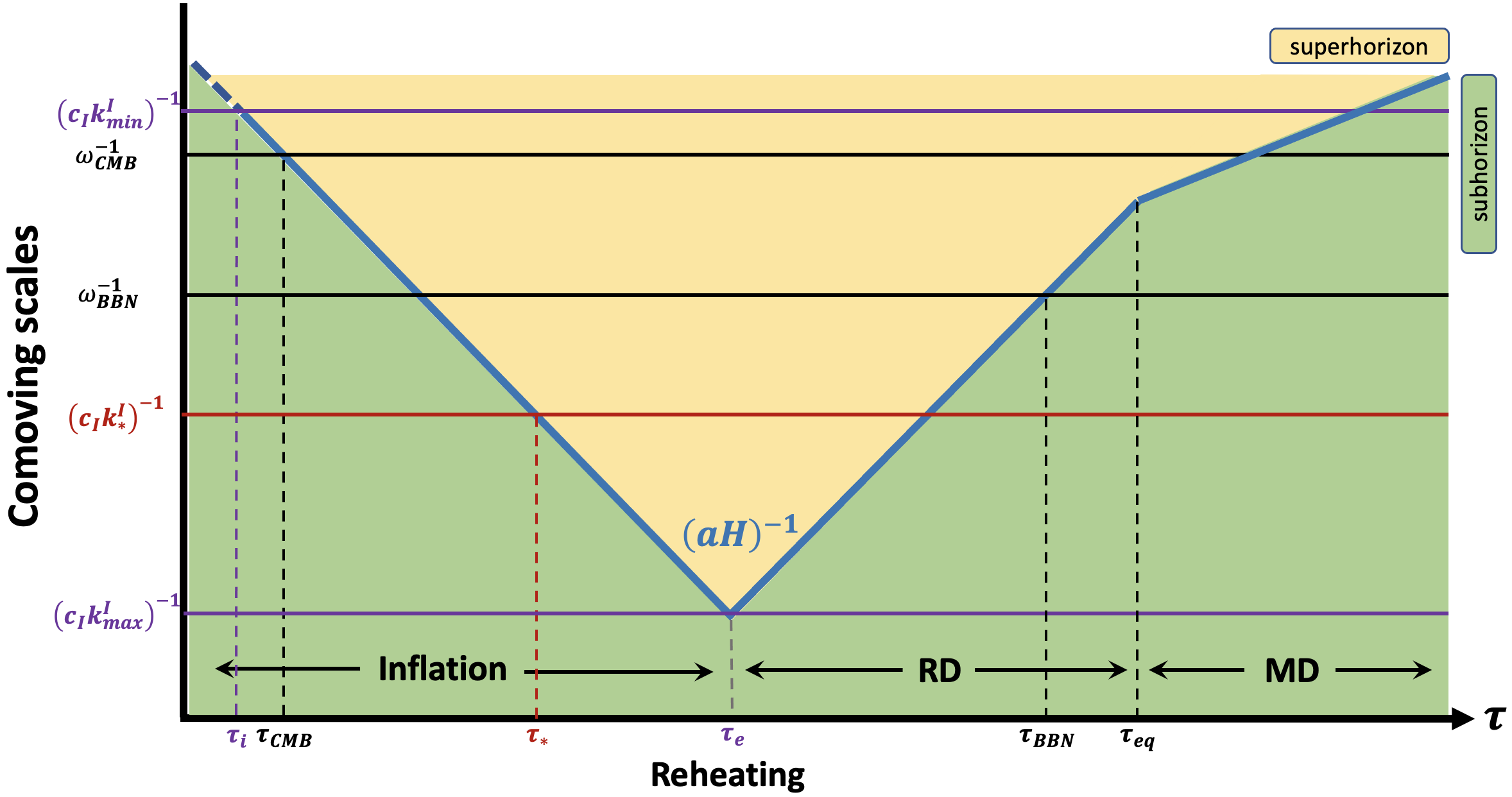}
	\caption{The light spin-2 particles with mass $m\ll{H}_{\rm inf}$ produce from tachyonic modes $k^I_{\min}\lesssim{k}\lesssim{k}^I_{\max}$. The modes $k^I_\ast<{k}\leq{k}^I_{\max}$ remain relativistic until BBN where $c_Ik^I_{\ast}$ corresponds to the scale/frequency that becomes non-relativistic at $\tau=\tau_{\rm BBN}$. All modes become non-relativistic before the time of matter and radiation equality $\tau=\tau_{\rm eq}$. The produced spin-2 particles dominate energy density of the universe around $\tau=\tau_{\rm eq}$ and play the role of DM. The CMB and BBN scales are roughly around $\omega_{\rm CMB}\sim{\cal O}\left(10^{-2}\right)\mbox{Mpc}^{-1}$ and $\omega_{\rm BBN}\sim10^{3}\mbox{Mpc}^{-1}$.}
	\label{fig-modes}
\end{figure}

The tachyonic modes defined by $\omega_I^2<0$ in \eqref{omega-inflation} are given by
\begin{align}\label{k-limits}
&c_Ik^I_{\min} = - \frac{\sqrt{2}}{\tau_{\rm i}} \,,
&c_Ik^I_{\max} = - \frac{\sqrt{2}}{\tau_{\rm e}} \,,
\end{align}
where $c_Ik^I_{\min}$ corresponds to the largest CMB scale which leaves the horizon at $\tau_{\rm i}$ and $c_Ik^I_{\max}$ corresponds to the smallest CMB mode which leaves the horizon at end of inflation $\tau=\tau_{\rm e}$ (see Fig.~\ref{fig-modes}). We are interested in the energy density at the end of inflation and we find
\begin{align}\label{modes-tachyonic}
&x^I_{\min} 
= \sqrt{2} \frac{\tau}{\tau_{\rm i}}\Big{|}_{\tau = \tau_{\rm e}}
= \sqrt{2}\, e^{-{\cal N}_{\rm e}} \,, 
&x^I_{\max} 
= \sqrt{2} \frac{\tau}{\tau_{\rm e}}\Big{|}_{\tau = \tau_{\rm e}}
= \sqrt{2} \,,
\end{align}
where ${\cal N}_{\rm e}=\ln(\tau_{\rm i}/\tau_{\rm e})$ is the total number of e-folds during inflation. Substituting \eqref{mode-function} in \eqref{energy-density} and performing integration for the tachyonic modes defined by \eqref{k-limits} or \eqref{modes-tachyonic}, it is straightforward to find
\begin{align}\label{ED-e}
\Omega_{\Sigma,{\rm e}} 
= {\cal C}_{\rm e}(c_I,{\cal N}_{\rm e})  
\left(\frac{H_{\rm e}}{2\pi\Mpl}\right)^2 \,,
\end{align}
where
\begin{align}\label{C-e}
{\cal C}_{\rm e}(c_I,{\cal N}_{\rm e}) 
= \left[
1 + \frac{4}{3} \ct^2 \left(1+{\cal N}_{\rm e}\right)
\right] 
\left(
\frac{1}{\cs^3} + \frac{2}{\cv^3} + \frac{2}{\ct^3} \right) \,,
\end{align}
in which we have neglected terms which are exponentially suppressed for ${\cal N}_{\rm e}={\cal O}(50-60)$.

In computing \eqref{ED-e}, we have taken the integration until the end of inflation but right before the deviation from quasi-dS background starts. Therefore, to be precise, $H_{\rm e}$ is not the Hubble parameter at the end of inflation but the Hubble parameter toward the end of inflation when the background is still quasi-dS and thus $H_{\rm e}\lesssim{H}_{\rm inf}$. However, we will assume an instantaneous reheating later and we identify $H_{\rm e}$ with the Hubble parameter at the end of reheating.

\subsection{Non-relativistic condition}

As an appropriate DM candidate, mass of the spin-2 field should be heavy enough to have a non-relativistic dust-like fluid. More precisely, conditions $m\gtrsim{H}$ and $m\gtrsim({c}_I{k})/a$ should meet for any excited modes $k^I_{\min}\leq{k}\leq{k}^I_{\max}$ before the time of matter and radiation equality.\footnote{We assume that $c_I$ and $m$ remain almost constant even after inflation. The time dependency of $c_I$ and $m$ are originated from the interaction with the inflaton field and, therefore, this assumption strongly depends on the forms of the interactions between the inflaton and the spin-2 field during whole period of inflation. In this regard, constancy of $c_I$ and $m$ should be more carefully considered when one deals with a particular model.} To guarantee the latter condition, we can only look at $m\gtrsim({c}_I{k}^I_{\max})/a$ since ${c}_I{k}^I_{\max}$ corresponds to the largest frequency (smallest scale) that becomes non-relativistic by the expansion of the universe. We thus need to compare two scales $H$ and $(c_I{k}^I_{\max})/a$. From \eqref{k-limits}, we find the following expression
\begin{align}\label{k-max}
c_I k^I_{\rm max} \simeq a_{\rm e}H_{\rm e} \,,
\end{align}
for the largest frequency that satisfies $\omega_I^2<0$ which gives
\begin{align}
\frac{c_I{k}^I_{\max}}{aH} = \frac{(aH)^{-1}}{(a_{\rm e}H_{\rm e})^{-1}} \,.
\end{align}
As shown in Fig.~\ref{fig-modes}, the comoving Hubble horizon $(aH)^{-1}$ is decreasing during inflation while it is increasing after inflation. Thus, $(c_I{k}^I_{\max})/a$ is always larger than $H$ and, therefore, we only need to make sure that $m\gtrsim(c_I{k}^I_{\max})/a$ satisfies before the time of matter and radiation equality. We define non-relativistic time $\tau=\tau_{\rm NR}$ at which all modes become non-relativistic. The corresponding scale factor is given by
\begin{align}\label{a-NR}
a_{\rm NR} \equiv \frac{c_I k^I_{\max}}{m} = \left(\frac{H_{\rm e}}{m}\right) a_{\rm e}\,,
\end{align}
which determines the time of complete transition from relativistic regime $\Omega_{\Sigma} \propto{a}^{-4}$ to non-relativistic regime $\Omega_{\Sigma} \propto{a}^{-3}$. The transition should happen during the radiation domination when $\Omega_{\Sigma}$ is a subdominant component. After $\tau=\tau_{\rm NR}$, $\Omega_{\Sigma}$ decays with less rate than radiation component and finally dominates the energy density of the universe at the time of matter and radiation equality $\tau=\tau_{\rm eq}$ and plays the role of DM (see Fig.~\ref{fig-modes}).

All modes have to become non-relativistic before the time of matter and radiation equality and, therefore, $\tau_{\rm NR}\lesssim{\tau}_{\rm eq}$ or $a_{\rm NR}\lesssim{a}_{\rm eq}$. Moreover, as we have already mentioned, the produced DM particles should have light mass $m\ll{H}_{\rm inf}$ as it is shown in Eq. \eqref{mass-light}. As $H_{\rm e}\lesssim{H}_{\rm inf}$, we consider the strong constraint of $m\ll{H}_{\rm e}$. These conditions together lead to the following allowed range for the mass
\begin{align}\label{mass-range}
&m_{\min} \lesssim m \ll m_{\max} \,;
&m_{\min} \equiv \left(\frac{a_{\rm e}}{a_{\rm eq}}\right) H_{\rm e} \,,
\hspace{.5cm}
m_{\max} \equiv H_{\rm e}\,.
\end{align}
Eq. \eqref{mass-range} represents the minimum condition which should be satisfied by mass of the spin-2 DM particles which are produced during inflation.

\section{Constraints on the model}\label{sec-constraints}

Before finding the DM relic today, we look at the theoretical and observational constraints on the parameters of our model in this section.

\subsection{Backreaction during inflation}\label{sec-backreaction}

The first constraint that may come in mind is that overproduction of spin-2 particles during inflation may destroy inflationary background which is known as the backreaction problem. This means that energy density of the spin-2 field should be subdominant compared with the energy density of the inflaton field. Ignoring interaction term in the Einstein equations \eqref{EEs}, which is subdominant in our weakly coupled EFT setup, the temporal component (the first Friedmann equation), can be written as $\Omega_\phi+\Omega_\Sigma=1$ where $\Omega_\phi$ is the fractional energy density of the inflaton. To avoid backreaction problem, the inflaton field should dominate the background energy density $\Omega_\phi\simeq1$ which implies
\begin{align}\label{backreaction-0}
& \Omega_\Sigma = \frac{1}{8} {\cal C} \, {\cal P}_{h}
=  \frac{1}{8} {\cal C}  \, r{\cal P}_{\cal R}\ll1 \,;
&{\cal P}_{h} = \frac{2H^2}{\pi^2\Mpl^2} \,,
\end{align}
where ${\cal P}_{h}$ is the power spectrum of the metric tensor perturbations and $r={\cal P}_{h}/{\cal P}_{\cal R}$ is the tensor-to-scalar ratio. At CMB scales $\omega_{\rm CMB}={\cal O}\left(10^{-2}\right)\mbox{Mpc}^{-1}$ (see Fig.~\ref{fig-modes}), cosmic microwave background (CMB) observations put constraints on the amplitude of the curvature perturbation power spectrum and tensor-to-scalar ratio as ${\cal P}_{\cal R}={\cal O}\left(10^{-9}\right)$ and $r<0.03$ \cite{Planck:2018jri}. This means $\Omega_{\Sigma,{\rm CMB}} < {\cal O}\left(10^{-12}\right){\cal C}$ and Eq. \eqref{backreaction-0} implies ${\cal C}_{\rm CMB} \ll {\cal O}\left(10^{12}\right)$. On the other hand, ${\cal C}_{\rm e}$ is the largest possible value of ${\cal C}$ since more and more modes become tachyonic toward the end of inflation. This accumulative effect on ${\cal C}$ corresponds to changing the value of number of e-folds and, therefore, ${\cal C}_{\rm e}$ can be larger than ${\cal C}$ at any moment at most by a factor proportional to the total number of e-folds. In this regard, ${\cal C}_{\rm e}>{\cal C}_{\rm CMB}$ and the strongest condition to avoid backreaction problem is
\begin{align}\label{backreaction-1}
&{\cal C}_{\rm e} \ll {\cal O}\left(10^{12}\right) \,;
&\mbox{backreaction constraint} \,.
\end{align}

It is also worth mentioning that the above condition is only based on the first Friedmann equation. Similar conditions may be concluded from other equations of motion. For instance, action $S_{\Sigma}$ defined in \eqref{action-free-general} depends on $\phi$ and, therefore, it contributes to the equation of motion of $\phi$ as a source. In an explicit model, one may need to check all background equations (see for instance \cite{Salehian:2020asa}).

\subsection{Isocurvature perturbations}

Even if backreaction condition \eqref{backreaction-1} satisfies and energy density of spin-2 field be very small compared to the energy density of inflaton, still it has a non-zero contribution to the total background energy density. Therefore, it gives a subdominant contribution to the curvature perturbation. One can then look for the so-called curvature/isocurvature decomposition \cite{Gordon:2000hv,Wands:2007bd,Lalak:2007vi,Langlois:2008mn} of the total energy-momentum tensor, which is defined in Eq. \eqref{EEs}, to find contribution of spin-2 field to the curvature/isocurvature perturbation. Here, however, we implement an intuitive but simpler approach based on the $\delta{\cal N}$ formalism \cite{Sasaki:1995aw,Lyth:2004gb,Wands:2000dp,Abolhasani:2019cqw}. 

As the spin-2 field is the only field other than inflaton and it is also subdominant field during inflation, isocurvature/entropy perturbation ${\cal S}$ should receive its dominant contribution from the spin-2 field perturbations. Following \cite{Salehian:2020asa}, we thus estimate it as
\begin{align}
{\cal S} 
\simeq \frac{\delta\rho_{\Sigma}}{\rho_{\Sigma}} 
=  \frac{\delta{\cal C}}{{\cal C}} \,,
\end{align}
where we have used Eq. \eqref{Omega-sigma} in the last step. As we will show, ${\cal C}$ is a function of time through its dependency on the number of e-folds ${\cal N}$ and also $c_I$. Since we have assumed that sound speeds change very slowly in time, we find
\begin{align}
{\cal S} 
\simeq \left( \frac{1}{{\cal C}} \frac{\partial{\cal C}}{\partial{\cal N}} \right) \delta{\cal N} 
= - \left(\frac{\partial\ln{\cal C}}{\partial{\cal N}}\right) {\cal R} \,,
\end{align}
where we have used the $\delta{\cal N}=-{\cal R}$ in the last step. In this respect, we find
\begin{align}
\beta \equiv \frac{{\cal P}_{\cal S}}{{\cal P}_{\cal R}} = \left(\frac{\partial\ln{\cal C}}{\partial{\cal N}}\right)^2 \,.
\end{align}

The constraint from isocurvature perturbations is $\beta_{\rm CMB}\lesssim 10^{-3}$ which holds for the CMB scales around $\omega\lesssim\omega_{\rm CMB}\sim{\cal O}\left(10^{-2}\right)\mbox{Mpc}^{-1}$ \cite{Planck:2018jri}. We thus need to compute ${\cal C}$ for the scales that become superhorizon at CMB scales between $\tau_{\rm i}$ and $\tau_{\rm CMB}$. $\tau_{\rm CMB}$ denotes the time that last CMB mode leaves the horizon (see Fig.~\ref{fig-modes}). It is straightforward to show that the corresponding tachyonic modes are determined by $c_Ik^I_{\min} = - \sqrt{2}/\tau_{\rm i}$ and $c_Ik^I_{\max} = - \sqrt{2}/\tau_{\rm CMB}$ which leads to
\begin{align}\label{modes-tachyonic-CMB}
&x^I_{\min} 
= \sqrt{2} \frac{\tau}{\tau_{\rm i}}\Big{|}_{\tau = \tau_{\rm CMB}}
= \sqrt{2} e^{-{\cal N}_{\rm CMB}} \,, 
&x^I_{\max} 
= \sqrt{2} \frac{\tau}{\tau_{\rm CMB}}\Big{|}_{\tau = \tau_{\rm CMB}}
= \sqrt{2} \,,
\end{align}
where ${\cal N}_{\rm CMB}=\ln(\tau_{\rm i}/\tau_{\rm CMB})=\ln(a_{\rm CMB}/a_{\rm i})$ denotes the number of e-folds that takes for all CMB modes to become superhorizon. We then find
\begin{align}\label{C}
{\cal C}_{\rm CMB}(c_I,{\cal N}_{\rm CMB}) 
= \left[
1 + \frac{4}{3} \ct^2 \left(1+{\cal N}_{\rm CMB}\right)
\right] 
\left(
\frac{1}{\cs^3} + \frac{2}{\cv^3} + \frac{2}{\ct^3} \right) \,,
\end{align}
where we have neglected terms which are exponentially suppressed for ${\cal N}_{\rm CMB}={\cal O}(10)$. The isocurvature constraint implies
\begin{align}
&\beta_{\rm CMB} \simeq \left[
\frac{\frac{4}{3}\ct^2}{1+\frac{4}{3}\ct^2\left(1+{\cal N}_{\rm CMB}\right)} 
\right]^2 
\lesssim 10^{-3} \,;
&\mbox{isocurvature constraint} \,.
\end{align}
For $c_t\ll\sqrt{3/4{\cal N}_{\rm CMB}}\simeq0.27$, $\beta_{\rm CMB}\simeq\left(4\ct^2/3\right)^2$ and $\beta_{\rm CMB}\lesssim10^{-3}$ can be very easily satisfied. For $\sqrt{3/4{\cal N}_{\rm CMB}}\ll\ct\lesssim1$, we find $\beta_{\rm CMB}\simeq{\cal N}_{\rm CMB}^{-2}={\cal O}\left(10^{-2}\right)$. Thus, the isocurvature constraint for $\ct={\cal O}(1)$ should be checked more systematically which is beyond the scope of this work. We, however, keep $\ct={\cal O}(1)$ as an option keeping in mind that the isocurvature constraint may or may not rule out this case. Note also that the isocurvature constraint can be completely relaxed by choosing different parameterization than \eqref{parametrization} (see for instance \cite{Firouzjahi:2021lov}).

\subsection{Big Bang nucleosynthesis}

All superhorizon modes $k^I_{\min}\leq{k}\leq{k}^I_{\max}$ excite and contribute to the energy density of DM in our scenario. These modes become non-relativistic at different times. Then, it is important to see which modes remain relativistic before the time of Big Bang nucleosynthesis (BBN) as those contribute to the effective number of relativistic degrees of freedom and, in this respect, we find constraints on the parameters of our model from BBN.

As $c_Ik^I_{\min}$ corresponds to the scale which leaves the horizon (or equivalently modes $k^I_{\min}$ leave their sound horizons) at CMB scales $c_Ik^I_{\min}\lesssim{\cal O}\left(10^{-2}\right)\mbox{Mpc}^{-1}$, we have $c_Ik^I_{\rm min}\ll\omega_{\rm BBN}={\cal O}\left(10^{3}\right)\mbox{Mpc}^{-1}$ where $\omega_{\rm BBN}=c_I k^I_{\rm BBN}$ denotes the BBN scale (see Fig.~\ref{fig-modes}). This means the modes $k^I_{\min}\leq{k}<{k}^I_{\rm BBN}$ leave the horizon sooner than ${k}^I_{\rm BBN}$ during inflation and re-enter the horizon after BBN during the radiation domination. Thus, the modes $k^I_{\min}\leq{k}<{k}^I_{\rm BBN}$ remain superhorizon until BBN and will not contribute to the effective number of relativistic degrees of freedom independent of when they become non-relativistic. Indeed, we will show that they all become non-relativistic before the time of BBN.

Now we look for the characteristic scale $c_Ik^I_{\ast}$ that becomes non-relativistic at the time of BBN
\begin{align}\label{k-star}
\frac{c_Ik^I_{\ast}}{m} = a_{\rm BBN} \, 
\quad
\Rightarrow
\quad
c_Ik^I_{\ast} = \left(\frac{m}{H_{\rm BBN}} \right) \omega_{\rm BBN} \,.
\end{align}
The ratio $m/H_{\rm BBN}$ determines whether $c_Ik^I_{\ast}$ is greater than $\omega_{\rm BBN}$ or not and we are going to find a rough estimation for it. Assuming instantaneous reheating, we find $\left(a_{\rm e}/a_{\rm eq}\right)=\left(T_{\rm eq}/T_{\rm r}\right)\left(g_{s\ast,{\rm eq}}/g_{s\ast,{\rm r}}\right)^{1/3}$ and $H_{\rm e}/H_{\rm BBN}=\left(g_{\ast,{\rm r}}/g_{\ast,{\rm BBN}}\right)^{1/2}\left(T_{\rm r}^2/T_{\rm BBN}^2\right)$ which give $m_{\min}/H_{\rm BBN}={\cal O}\left(T_{\rm eq}T_{\rm r}/T_{\rm BBN}^2\right)$ where $m_{\min}$ is defined in \eqref{mass-range}. For $T_{\rm BBN}=10^{-1}\mbox{MeV}$ and $T_{\rm eq}=10^{-6}\mbox{MeV}$, any values of $T_{\rm r}\gtrsim10\mbox{GeV}$ implies $m_{\min}> H_{\rm BBN}$. As we deal with $T_{\rm r}={\cal O}\left(10^{10}-10^{15}\right)\mbox{GeV}$, we find $m_{\min}\gg{H}_{\rm BBN}$ which implies $m\gg{H}_{\rm BBN}$.\footnote{For larger values of BBN temperature, we need larger values of $T_{\rm r}$ to guaranty $m_{\min}\gg{H_{\rm BBN}}$. For example, $T_{\rm r}>10^{4}\mbox{GeV}$ is needed for $T_{\rm BBN}=4-5\mbox{MeV}$ \cite{Hasegawa:2019jsa}. However, we deal with $T_{\rm r}={\cal O}\left(10^{10}-10^{15}\right)\mbox{GeV}$ in this paper, with which it is easy to satisfy $m_{\min}\gg{H_{\rm BBN}}$.} From \eqref{k-star} we then conclude $c_I k^I_\ast\gg\omega_{\rm BBN}$ or equivalently $k^I_\ast\gg{k}^I_{\rm BBN}$. From this result, we immediately find that all modes $k^I_{\rm BBN}\leq{k}\leq{k}^I_\ast$ become non-relativistic before BBN while $k^I_\ast<{k}\leq{k}^I_{\max}$ remain relativistic before BBN. Depending on the mass, either $k^I_\ast\ll{k}^I_{\max}$ or $k^I_\ast\sim{k}^I_{\max}$ may be possible. Let us make an estimation of the ratio $k^I_\ast/{k}^I_{\max}$. From \eqref{k-star} and \eqref{k-max} we find $k^I_\ast/{k}^I_{\max}=(a_{\rm BBN}/a_{\rm e})(m/H_{\rm e})$. The lower limit for the mass in \eqref{mass-range} then implies $k^I_\ast/{k}^I_{\max}\gtrsim{a}_{\rm BBN}/a_{\rm eq}$ which after using $\left(a_{\rm BBN}/a_{\rm eq}\right)=\left(T_{\rm eq}/T_{\rm BBN}\right)\left(g_{s\ast,{\rm eq}}/g_{s\ast,{\rm BBN}}\right)^{1/3}$ leads to
\begin{align}\label{k-star-range}
{\cal O}\left(\frac{T_{\rm eq}}{T_{\rm BBN}}\right) \lesssim \frac{k^I_\ast}{{k}^I_{\max}} \lesssim 1 \,,
\end{align}
where the upper limit corresponds to $k^I_\ast\sim{k}^I_{\max}$ when all modes become relativistic before the time of BBN. For $T_{\rm BBN}=10^{-1}\mbox{MeV}$ and $T_{\rm eq}=10^{-6}\mbox{MeV}$ we find $k^I_\ast\sim{\cal O}\left(10^{-5}\right){k}^I_{\max}$. This simple estimation shows that $k^I_\ast\ll{k}^I_{\max}$ is possible and, therefore, we will get a nontrivial bound from BBN. 

We have shown that the modes $k^I_\ast<{k}\leq{k}^I_{\max}$, where $k^I_\ast$ and $k^I_{\max}$ are given by \eqref{k-star} and \eqref{k-max} respectively, can remain relativistic until BBN for a reasonable mass range. These modes will contribute to the number of effective relativistic degrees of freedom from which we will find a BBN constraint on the energy density. We need to find contribution of modes $k^I_\ast<{k}\leq{k}^I_{\max}$ to the energy density \eqref{ED-e}. In order to do so, we need to find at which time $k^I_\ast$ become tachyonic. The condition $\omega_I^2<0$ in \eqref{omega-inflation} can be achieved for $\tau\geq\tau_\ast$ where $\tau_\ast=-\sqrt{2}/(c_I k^I_\ast)$ such that
\begin{align}\label{modes-tachyonic-star}
&x^I_{\ast\min} 
= \sqrt{2} \frac{\tau}{\tau_\ast}\Big{|}_{\tau = \tau_{\rm e}}
= \sqrt{2} e^{-\Delta{\cal N}} \,,
&x^I_{\max} 
= \sqrt{2} \frac{\tau}{\tau_{\rm e}}\Big{|}_{\tau = \tau_{\rm e}}
= \sqrt{2} \,,
\end{align}
where $\Delta{\cal N}=\ln(\tau_{\rm e}/\tau_\ast)={\cal N}_{\rm e}-{\cal N}_{\ast}$ is the number of e-folds from the time $\tau_\ast$ when $k^I_\ast$ becomes tachyonic until the end of inflation when $k_{\max}$ becomes tachyonic (see Fig.~\ref{fig-modes}).

Using solution for the mode function \eqref{mode-function} in \eqref{energy-density} and performing integration for the tachyonic modes $k^I_{\ast}\leq{k}\leq{k}^I_{\max}$ with upper and lower limits of the integral given by \eqref{modes-tachyonic-star}, we find
\begin{align}\label{ED-e-star}
\Omega^\ast_{\Sigma,{\rm e}} 
= {\cal C}^\ast_{\rm e}(c_I,\Delta{\cal N})  
\left(\frac{H_{\rm e}}{2\pi\Mpl}\right)^2 \,,
\end{align}
where we have defined
\begin{align}
{\cal C}^\ast_{\rm e}(c_I,\Delta{\cal N}) 
= 
\left[1+\frac{4}{3}\ct^2\left(1+\Delta\mathcal{N}\right)\right] \left(
\frac{1}{\cs^3} + \frac{2}{\cv^3} + \frac{2}{\ct^3} \right) \,,
\end{align}
in which we have neglected the terms which are exponentially suppressed in $\Delta{\cal N}$. 

In \eqref{ED-e-star}, we have computed contribution of the modes which remain relativistic until BBN to the energy density. As we have already mentioned, these modes contribute to the effective number of relativistic degrees of freedom. We thus need to keep track the time evolution of ${\cal C}^\ast_{\rm e}(c_I,\Delta{\cal N})$ in \eqref{ED-e-star} until the time of BBN. Since we have assumed an instantaneous reheating, $\Omega^\ast_{\Sigma,{\rm e}} $ scales like radiation $a^{-4}$ and we only need to take into account changes in effective number of relativistic degrees of freedom. From conservation of the entropy we find $\Omega^\ast_{\Sigma,{\rm BBN}}=\Omega^\ast_{\Sigma,{\rm e}} \left(g_{\ast,{\rm BBN}}/g_{\ast,{\rm r}}\right) \left(g_{s\ast,{\rm BBN}}/g_{s\ast,{\rm r}}\right)^{4/3}$. Since the effective number of relativistic degrees of freedom decreases in time, we have $\Omega^\ast_{\Sigma,{\rm e}}>\Omega^\ast_{\Sigma,{\rm BBN}}$. Thus, taking into account the effect of changes in the effective number of relativistic degrees of freedom will make the BBN constraint weaker. In this regard, $\Delta{N}_{{\rm eff}, \max} = \frac{8}{7}\left(\frac{11}{4}\right)^{4/3} \Omega^\ast_{\Sigma,{\rm e}}$ and we find
\begin{align}
\Omega^\ast_{\Sigma,{\rm e}} =
\frac{7}{8}\left(\frac{4}{11}\right)^{4/3} 
\Delta{N}_{{\rm eff}, \max} < 0.068
\,,
\end{align}
where we have used $\Delta{N}_{\rm eff}<0.3$ \cite{Planck:2018vyg}. Rewriting $\Omega_{\Sigma,{\rm e}}$ in terms of $\Omega^\ast_{\Sigma,{\rm e}}$ we find 
 \begin{align}\label{BBN-constraint-0}
\left(\frac{{\cal C}^\ast_{\rm e}}{{\cal C}_{\rm e}}\right) \Omega_{\Sigma,{\rm e}} < 0.068 \,.
 \end{align}
Note that ${\cal C}^\ast_{\rm e}/{\cal C}_{\rm e}\leq1$ since $\Delta{\cal N}\leq{\cal N}_{\rm e}$. Thus, the strongest constraint from BBN corresponds to $\Omega_{\Sigma,{\rm e}} < 0.068$.\footnote{To be more precise, ${\cal C}^\ast_{\rm e}\sim{\cal C}_{\rm e}$ needs $k^I_\ast\sim{k}^I_{\min}$ which means all modes remain relativistic until BBN. However, from Eq. \eqref{k-star-range}, we see that the minimum value of $k^I_\ast$ is given by $k^I_\ast={\cal O}\left(T_{\rm eq}/T_{\rm BBN}\right){k}^I_{\max}={\cal O}\left(10^{-5}\right){k}^I_{\max}\gg{k}^I_{\min}$ since ${k}^I_{\max}\sim\exp({\cal N}_{\rm e}){k}^I_{\min}\sim{\cal O}\left(10^{26}\right){k}^I_{\min}$ for ${\cal N}_{\rm e}={\cal O}(60)$. This simple estimation shows that ${\cal C}^\ast_{\rm e}<{\cal C}_{\rm e}$. Nevertheless, the difference between ${\cal C}^\ast_{\rm e}$ and ${\cal C}_{\rm e}$ is proportional to $\ln(k^I_\ast/{k})$ which can change the result at most as $\ln(k^I_\ast/{k}^I_{\min})\sim({\cal N}_{\rm e}-\Delta{\cal N})$. Thus, the constraint \eqref{BBN-constraint} will be weaken by almost ten/hundred order of magnitude.} In this case, as we have estimated $\Omega_{\Sigma,{\rm e}} 
\simeq {\cal O}(10^{-12})\, {\cal C}_{\rm e}$ in Sec. \ref{sec-backreaction}, Eq. \eqref{BBN-constraint-0} implies
\begin{align}\label{BBN-constraint}
&{\cal C}_{\rm e} < 6.8 \times 10^{10} \,;
&\mbox{BBN constraint} \,.
\end{align}
On the other hand $ {\cal C}^\ast_{\rm e}/{\cal C}_{\rm e}\to0$ for $k^I_\ast \sim {k}^I_{\max}$. In that case, all modes become non-relativistic before BBN and there will be no BBN constraint on the model.

The BBN constraint \eqref{BBN-constraint} is stronger than backreaction constraint \eqref{backreaction-1}. Indeed BBN constraint is the strongest constraint that we find on ${\cal C}_{\rm e}$.

\subsection{Stability}

Finally, we discuss stability conditions which also puts constraints on the parameters of the model. 

First of all, action \eqref{action} should be free of ghost and gradient instabilities. The ghost stability is guarantied by our canonical choice for the coefficients $A, B, C,$ in \eqref{action} while it would give nontrivial conditions on $A, B, C$ for general case given by action \eqref{action-free-Sigma-p}. For instance, in the absence of any interaction with the inflaton field, it is not possible to satisfy stability conditions for all scalar, vector, and tensor modes for $m^2\leq2{H}^2$ leading to the Higuchi bound \eqref{Higuchi-bound}. The gradient stability for the scalar, vector, and tensor modes defined by actions \eqref{action-free-S}, \eqref{action-free-V}, and \eqref{action-free-T} need
\begin{align}\label{gradient-stability}
&\cs^2, \cv^2, \ct^2 > 0 \,.
\end{align}
Indeed, with our parameterization \eqref{parametrization}, as it can be clearly seen in Eq. \eqref{cv}, $\cv$ is not an independent parameter and $\cs^2>0$ and $\ct^2>0$ automatically imply $\cv^2>0$. Note, however, that conditions \eqref{gradient-stability} give quite nontrivial constraints in general.

Secondly, as it is shown in Eq. \eqref{action-int-general}, there are direct interactions between the spin-2 field, inflaton and gravity. Based on the EFT derivative expansion, in our weakly coupled EFT setup, these interactions are suppressed compared to the leading terms included in free action \eqref{action}. Nevertheless, they provide corrections to the spectra of curvature perturbations and primordial gravitational waves even at the linear level: the scalar mode $s$ and tensor modes $t_{ij}$ source linear equation of motions of curvature perturbations and metric tensor perturbations respectively \cite{Bordin:2018pca,Bordin:2019tyb,Iacconi:2019vgc,Iacconi:2020yxn}. Similarly, there will be corrections to the power spectra and also energy density from the interaction term \eqref{action-int-general} which are negligible compared to the contributions from the free actions as far as the setup is weakly coupled. That is why we have ignored the effects of the interaction term \eqref{action-int-general} in computing energy density of the spin-2 field. However, if the amplitude of spin-2 particles is enhanced, interactions may become important and possibly give rise to a perturbative decay of the spin-2 particles to inflaton and/or gravitons. Indeed, in order to produce enough DM, large amplitude of perturbations is not needed at all since we are interested in accumulated energy density of spin-2 particles. To see this fact, let us make an estimation of the amplitude of the excited perturbations of the spin-2 field. The power spectrum for the normalized field defined in \eqref{XI} is given by $\langle{X}_{\bf k}^I X_{\bf q}^J \rangle = P_I(k,\tau) (2\pi)^3\delta^{IJ} \delta({\bf k}-{\bf q})$ which gives $P_I(k,\tau)=\big| X_{k}^I \big|^2$. Then, defining the dimensionless power spectra for the canonical fields as usual ${\cal P}_{\rm s}=\left(k^3/2\pi^2\right)P_{S}/a^2$, ${\cal P}_{\rm v}=\left(k^3/2\pi^2\right)(P_{V^{+}}+P_{V^{\times}})/a^2$, and ${\cal P}_{\rm t}=\left(k^3/2\pi^2\right)(P_{T^{+}}+P_{T^{\times}})/a^2$, we find
\begin{align}\label{PS-sigma-svt}
&{\cal P}_{\rm s} = \left(\frac{H_{\rm inf}}{2\pi\Mpl}\right)^2 \frac{1}{\cs^3} \,,
&&{\cal P}_{\rm v} = \left(\frac{H_{\rm inf}}{2\pi\Mpl}\right)^2 \frac{2}{\cv^3} \,,
&&{\cal P}_{\rm t} = \left(\frac{H_{\rm inf}}{2\pi\Mpl}\right)^2 \frac{2}{\ct^3} \,.
\end{align}

Considering a tachyonic window defined by some $k^I_{\min}$ and $k^I_{\max}$, we find the following relation between the energy density and power spectrum of the spin-2 field during whole inflationary period
\begin{align}\label{PS-sigma-0}
&\Omega_\Sigma 
= \left[
1 + \frac{4}{3} \ct^2 \left(1+{\cal N}\right)
\right] {\cal P}_\Sigma \,;
&{\cal P}_\Sigma \equiv {\cal P}_{\rm s} + {\cal P}_{\rm v} + {\cal P}_{\rm t} \,,
\end{align}
where ${\cal N}=\ln(k^I_{\max}/k^I_{\min})$. For \eqref{k-limits}, we get ${\cal N}={\cal N}_{\rm e}$ while we do not fix the value of number of e-folds and it can be any period during inflation. Note that here we deal with completely scale-invariant power spectra such that power spectra simply represent amplitudes. In this respect, ${\cal P}_\Sigma$ corresponds to the amplitude of the power spectrum of the spin-2 field.

Let us rewrite the power spectra of the spin-2 field in terms of the power spectrum of the curvature perturbations. Using \eqref{backreaction-0} in \eqref{PS-sigma-0} yields 
\begin{align}\label{PS-sigma}
{\cal P}_\Sigma = \frac{1}{8}{\cal C} \left[
1 + \frac{4}{3} \ct^2 \left(1+{\cal N}\right)
\right]^{-1} \, r{\cal P}_{\cal R} \,.
\end{align}
At CMB scales $\omega_{\rm CMB}\sim{\cal O}\left(10^{-2}\right)\mbox{Mpc}^{-1}$, ${\cal P}_{\cal R}={\cal O}(10^{-9})$ and $r<0.03$ and ${\cal P}_{\Sigma,{\rm CMB}}\ll1$ implies ${\cal C}_{\rm CMB} \ll {\cal O}\left(10^{12}\right) \left[1 + \frac{4}{3} \ct^2 \left(1+{\cal N}_{\rm CMB}\right)\right]$. Assuming scale-invariant power spectrum for curvature perturbations and taking into account the fact that ${\cal C}_{\rm e}>{\cal C}_{\rm CMB}$, we find
\begin{align}\label{constraint-stability}
&{\cal C}_{\rm e} \ll {\cal O}\left(10^{12}\right) \left[
1 + \frac{4}{3} \ct^2 \left(1+{\cal N}_{\rm e}\right)
\right] \,;
&\mbox{stability constraint} \,.
\end{align}
The above constraint is always weaker than the BBN constraint \eqref{BBN-constraint}. Depending on the value of $\ct$, it can be either at the same strength or weaker than the backreaction constraint \eqref{backreaction-1}.

Note that result \eqref{constraint-stability} holds only when both power spectrum of curvature perturbations and spin-2 field are scale-invariant. The first is an assumption in our setup while the latter is originated from our choices for EFT parameters $A,B,C,D$ in action \eqref{action}. For instance, by choosing different parametrization than \eqref{parametrization}, one may construct a model with a scale-dependent power spectrum that gets large amplification such that ${\cal P}_{\Sigma}\gg\Omega_\Sigma$ for some scale(s) while ${\cal P}_{\Sigma}\ll1$ and $\Omega_\Sigma\ll1$ (see for instance \cite{Gorji:2023ziy}). In that case, interactions in \eqref{action-int-general} may become important for the scale(s) at which the power spectrum is enhanced. Very interestingly, since spin-2 field provides tensor perturbations, the interaction term $\kappa$ in \eqref{action-int-general} provides a linear source for both metric vector and metric tensor perturbations. In the case of metric tensor perturbations, enhanced power spectrum of spin-2 field leads to the significant productions of stochastic gravitational waves \cite{Gorji:2023ziy}. It is then interesting to construct a spin-2 DM model with scale-dependent power spectrum as it will have a distinct signature on the spectrum of gravitational waves. We leave this possibility for a future work. 

It is worth mentioning that we have assumed that: i) the spin-2 field is the only field other than inflaton that exists during inflation, ii) it lives in the dark sector of the universe and it is decoupled from the standard model particles. These assumptions can be relaxed and one may assume direct interactions between the spin-2 field and other fields during inflation and after inflation. Then, assuming that the produced spin-2 DM particles do not decay (at least by the time scales of order of the age of the universe), one can find constraints on the corresponding couplings.

\section{DM relic density today}\label{sec-DM-relic}

Having found the allowed mass range and also constraints on the model, we find the relic density of DM today in this section.

In order to find energy density of DM today, we need to keep track the time evolution of the energy density of the produced spin-2 particles at the end of inflation \eqref{ED-e} until now. From $\tau=\tau_{\rm e}$ until $\tau=\tau_{\rm NR}$, energy density of spin-2 field scales as $a^{-4}$ and after $\tau=\tau_{\rm NR}$ it scales as $a^{-3}$. We thus find $\rho_{{\rm DM},0}=\left(a_{\rm e}/a_0\right)^3\left(a_{\rm e}/a_{\rm NR}\right)\rho_{\Sigma,{\rm e}}$ where $\rho_{{\rm DM},0}$ is the energy density of DM particles today.\footnote{More precisely, the mode $k_{\max}$ becomes non-relativistic at $\tau=\tau_{\rm NR}$ and all other modes become non-relativistic earlier. Had we taken into account this, the value of $\rho_{{\rm DM},0}$ would increase. Thus, by assuming that all modes become non-relativistic at $\tau=\tau_{\rm NR}$, we are estimating the minimum value of $\rho_{{\rm DM},0}$.} To keep the setup simple and theoretically under control, we assume instantaneous reheating
\begin{align}\label{T-reheating}
3 \Mpl^2 H_{\rm e}^2 = \frac{\pi^2}{30} g_{\ast,{\rm r}} T_{\rm r}^4 \,.
\end{align}
Then, from conservation of the entropy we find $\left(a_{\rm e}/a_0\right)^3=\left(s_0/s_{\rm e}\right)=\left(T_0/T_{\rm r}\right)^3\left(g_{s\ast,0}/g_{s\ast,{\rm r}}\right)$. Using this result, we find the following simple expression for the DM relic density today
\begin{align}\label{relic-DM-0}
\Omega_{\rm DM,{0}} 
&= \left( \frac{g_{\ast,{\rm r}} g_{s\ast,0}}{g_{\ast,0}g_{s\ast,{\rm r}}} \right) 
\left(\frac{T_{\rm r}}{T_0}\right)
\left(\frac{a_{\rm e}}{a_{\rm NR}}\right)
\Omega_{\Sigma,{\rm e}} 
= {\cal O}\left(10^{25}\right) \left(\frac{T_{\rm r}}{10^{12}\,{\mbox{GeV}}}\right)
\left(\frac{m}{H_{\rm e}}\right)
\Omega_{\Sigma,{\rm e}}
\,,
\end{align} 
where we have used Eq. \eqref{a-NR} in the second step. In the above result, $T_0\simeq10^{-13}\mbox{GeV}$ is the current temperature of the universe and we have also used $g_{s\ast,0}=3.91$, $g_{\ast,{\rm r}}=106.75=g_{s\ast,{\rm r}}$, $g_{\ast,0}=3.36$. Substituting \eqref{ED-e} in \eqref{relic-DM-0} and using \eqref{T-reheating}, we find
\begin{align}\label{relic-DM-0-f}
\Omega_{\rm DM,{0}} 
&= {\cal O}\left(10^{-16}\right) 
{\cal C}_{\rm e}(c_I,{\cal N}_{\rm e})
\left(\frac{T_{\rm r}}{10^{12}\,{\mbox{GeV}}}\right)^3 
\left(\frac{m}{1\mbox{eV}}\right)
\,.
\end{align}

The relic \eqref{relic-DM-0-f} depends on $c_I$, ${\cal N}_{\rm e}$, $T_{\rm r}$, and $m$. Dependency on $c_I$ and ${\cal N}_{\rm e}$ is characterized by ${\cal C}_{\rm e}(c_I,{\cal N}_{\rm e})$. To make the setup simple, we make order estimation of ${\cal C}_{\rm e}(c_I,{\cal N}_{\rm e})$ and leave $T_{\rm r}$ and $m$ as free parameters of the model. But, before that, we clarify possible ranges for the other two parameters $T_{\rm r}$ and $m$. 

It is important to note that assuming instantaneous reheating \eqref{T-reheating} strongly restricts our choices for $T_{\rm r}$ and $m$. First, fixing the value of $H_{\rm e}$ immediately fixes the reheat temperature $T_{\rm r}$. The parameter $H_{\rm e}$ is the Hubble parameter at the end of inflation. From the CMB observations we have ${\cal P}_{\cal R}={\cal O}\left(10^{-9}\right)$ and $r<0.03$ such that $H_{\rm e} \sim H_{\rm inf} \lesssim {\cal O}\left(10^{12}\right) \mbox{GeV}$ for $r={\cal O}\left(10^{-4}\right)$. We consider  $H_{\rm e} \sim H_{\rm inf}={\cal O}\left(10^{4}-10^{12}\right) \mbox{GeV}$ corresponds to $r={\cal O}\left(10^{-4}-10^{-22}\right)$. From the instantaneous reheating \eqref{T-reheating} we find that this range corresponds to the reheat temperatures $T_{\rm r}={\cal O}\left(10^{10}-10^{15}\right)\mbox{GeV}$. Secondly, using the instantaneous reheating \eqref{T-reheating} and $\left(a_{\rm e}/a_{\rm eq}\right)=\left(T_{\rm eq}/T_{\rm r}\right)\left(g_{s\ast,{\rm eq}}/g_{s\ast,{\rm r}}\right)^{1/3}$ with $g_{s\ast,{\rm eq}}\lesssim3.91$, Eq. \eqref{mass-range} can be rewritten as follows
\begin{align}\label{mass-range-TR}
{\cal O}\left(10^{-6}\right) \left(\frac{T_{\rm r}}{10^{12}\,{\mbox{GeV}}}\right)
\lesssim \frac{m}{1\mbox{eV}} \ll
{\cal O}\left(10^{15}\right) \left(\frac{T_{\rm r}}{10^{12}\,{\mbox{GeV}}}\right)^2 \,.
\end{align}

Now, we turn to the order estimation of the combination ${\cal C}_{\rm e}(c_I,{\cal N}_{\rm e})$. Looking at different regimes for $c_I$, \eqref{C-e} simplifies to 
\begin{equation}
{\cal C}_{\rm e}(c_I,{\cal N}_{\rm e}) = 
\begin{cases}
\begin{aligned}
&\frac{4}{3}\ct^2{\cal N}_{\rm e} \left(
\frac{1}{\cs^3} + \frac{2}{\cv^3} + \frac{2}{\ct^3} \right) 
& \cs,\cv,\ct={\cal O}(1) 
&& \mbox{\bf Case I}
\\
&\frac{2}{\ct} \left( \frac{1}{\ct^2}  + \frac{4}{3} {\cal N}_{\rm e} \right)
& \cs,\cv={\cal O}(1) , \,\,\,\,\,\, \ct\ll1 
&& \mbox{\bf Case II}
\\
&\frac{4}{3}\ct^2{\cal N}_{\rm e} \left(
\frac{1}{\cs^3} + \frac{2}{\cv^3} \right) 
& \cs,\cv\ll1, \,\,\,\,\, \ct={\cal O}(1) 
&& \mbox{\bf Case III}
\\
& \left[
1 + \frac{4}{3} \ct^2 \left(1+{\cal N}_{\rm e}\right)
\right] 
\left(
\frac{1}{\cs^3} + \frac{2}{\cv^3} + \frac{2}{\ct^3} \right)
& \cs,\cv,\ct\ll1 
&& \mbox{\bf Case IV}
\end{aligned}
\end{cases}
\end{equation}
The strongest constraint on ${\cal C}_{\rm e}$ is given by Eq. \eqref{BBN-constraint} which is provided by BBN. In order to estimate the order of magnitude of ${\cal C}_{\rm e}$, we should look for reasonable values of $c_I$ that do not violate the BBN bound. 

\begin{figure}[htbp!]
	\centering
	\includegraphics[width=.65 \columnwidth]{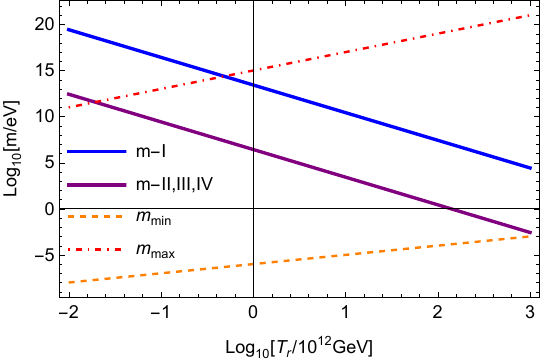}
	\caption{The upper blue solid curve shows mass of DM particles for case I as a function of reheat temperature given by \eqref{relic-DM-0-f} for $\Omega_{\rm DM,{0}}=0.27$. The lower purple solid curve corresponds to cases II, III, and IV. The red dot-dashed and orange dashed curves correspond to the upper and lower limits of mass given in \eqref{mass-range-TR} respectively. Any value of mass on the blue and purple solid curves which lies between the red dot-dashed and orange dashed curves can give the whole DM in the allowable mass range.}
\label{fig-1} 
\end{figure}

{\bf Case I}: In this case, we find ${\cal C}_{\rm e}={\cal O}(10^2)$ for ${\cal N}_{\rm e}={\cal O}(50-60)$ which is perfectly within the bound \eqref{BBN-constraint} imposed by BBN. For $\Omega_{\rm DM,{0}}=0.27$, we have plotted mass as a function of reheat temperature in Fig.~\ref{fig-1}. The plot is restricted to mass range determined by \eqref{mass-range-TR}. We will find that the following mass ranges can give the whole DM
\begin{align}
10\, \mbox{keV} \lesssim {m} \ll 10^{5}\, \mbox{GeV} \,,
\end{align} 
for reheat temperature $T_{\rm r}={\cal O}\left(10^{12}-10^{15}\right)\mbox{GeV}$. The minimum value for the mass in the above result is fixed by the maximum value of the reheat temperature $T_{\rm r}=10^{15}\mbox{GeV}$ while the maximum value of mass is set by $m_{\max}$ in \eqref{mass-range-TR} which is nothing but the value of Hubble parameter at the end of inflation $H_{\rm e}$. The latter is then a direct consequence of the instantaneous reheating \eqref{T-reheating}.\\

{\bf Case II}: For this case, depending on whether $\ct\ll\sqrt{3/4{\cal N}_{\rm e}}$ or $\ct\gg\sqrt{3/4{\cal N}_{\rm e}}$ we may have two possibilities. However, $\sqrt{3/4{\cal N}_{\rm e}}={\cal O}(10^{-1})$ for typical values ${\cal N}_{\rm e}={\cal O}(50-60)$. Therefore, the latter case is almost similar to the case I and we, therefore, do not consider it separately. For $\ct\ll\sqrt{3/4{\cal N}_{\rm e}}={\cal O}(10^{-1})$, we find ${\cal C}_{\rm e}\approx2\ct^{-3}$ which is independent of the number of e-folds. Then, we get large value of ${\cal C}_{\rm e}={\cal O}(10^9)$ for $\ct={\cal O}\left(10^{-3}\right)$. Note that smaller values $\ct<{\cal O}\left(10^{-3}\right)$ violate BBN bound \eqref{BBN-constraint}. For ${\cal C}_{\rm e}(c_I,{\cal N}_{\rm e})={\cal O}(10^9)$, from Fig.~\ref{fig-1}, we find that the following mass range
\begin{align}\label{mass-II}
10^{-3}\, \mbox{eV} \lesssim {m} \ll 10^{2}\, \mbox{GeV} \,,
\end{align}
with reheat temperatures $T_{\rm r}={\cal O}\left(10^{10}-10^{15}\right)\mbox{GeV}$
can give the whole DM.\\

{\bf Case III}: In this case, we find ${\cal C}_{\rm e}={\cal O}(10^9)$ for ${\cal N}_{\rm e}={\cal O}(50-60)$ and $\cs,\cv\simeq{\cal O}\left(10^{-2}\right)$. Smaller sound speeds $\cs,\cv<{\cal O}\left(10^{-2}\right)$ would violate BBN bound \eqref{BBN-constraint}. Fig.~\ref{fig-1} then shows that the mass range \eqref{mass-II} with reheat temperatures $T_{\rm r}={\cal O}\left(10^{11}-10^{15}\right)\mbox{GeV}$ can provide the whole DM energy density.\\

{\bf Case IV}: Similar to case II, we only consider $\ct\ll\sqrt{3/4{\cal N}_{\rm e}}$. We then find ${\cal C}_{\rm e}\approx\min\left[\cs,\cv,\ct\right]^{-3}$ which gives ${\cal C}_{\rm e}={\cal O}(10^9)$ for $\min\left[\cs,\cv,\ct\right]={\cal O}\left(10^{-3}\right)$. Using this estimation, from Fig.~\ref{fig-1}, we find that the mass range \eqref{mass-II} with reheat temperatures $T_{\rm r}={\cal O}\left(10^{11}-10^{15}\right)\mbox{GeV}$ can give the whole DM.\\

Thus, for reasonable values of $c_I$ which satisfy all constraints (or the strongest constraint \eqref{BBN-constraint} from BBN) and acceptable reheat temperatures, the mass of DM particles roughly lies in the wide range of $10^{-3}\, \mbox{eV} \lesssim {m} \ll 10^{5}\, \mbox{GeV}$. For $c_I={\cal O}(1)$, the possible minimum mass is $m=10\, \mbox{keV}$ while smaller values $10^{-3}\, \mbox{eV} \lesssim {m} \ll 10\,\mbox{keV}$ can be achieved for $c_I\ll1$. Moreover, it is not possible to produce the whole DM in our model for $T_{\rm r}\lesssim10^{10}\mbox{GeV}$. This means $T_{\rm r}\gtrsim 10^{10}\mbox{GeV}$ is needed for our scenario to work. Of course, this depends very much on our simplification by assuming simultaneous reheating \eqref{T-reheating} since this bound directly comes from the upper limit in \eqref{mass-range-TR} (see Fig.~\ref{fig-1}). Moreover, the mass range can be widened by choosing parametrization different than \eqref{parametrization} for EFT coefficients. We leave this possibility for future studies.

\section{Summary}\label{summary}

The seed of DM particles can be produced from light spectator fields during inflation through a similar mechanism that the seed of the observed large scale structures are generated by the inflaton field. The accumulated energy density of excited modes of the spectator fields, which is subdominant during inflation, dominates energy density of the universe around the time of matter and radiation equality and plays the role of DM. Recently, this idea is applied to the scalar and vector spectator fields. In a typical situation, spectator fields should have small mass (compared with the Hubble parameter during inflation) to allow for an efficient particle production process. In the case of spin-2 field, the so-called Higuchi bound \eqref{Higuchi-bound} seems to prevent excitation of such light modes. The previous studies on the production of spin-2 DM particles during inflation are then restricted to spectator fields with heavy masses which respect Higuchi bound \eqref{Higuchi-bound}. Indeed, the Higuchi bound is a direct consequence of the isometries of the dS space and since the spacetime is quasi-dS during inflation, one may expect that the deviation from the Higuchi bound should be very small. However, if there is a sizable interaction between the spin-2 field and inflaton field, (part) of the isometries of dS breaks and the Higuchi bound can be relaxed. We have used this possibility to construct a DM model in which spin-2 particles, as a source of DM, are produced during inflation through interactions with inflaton and also gravity. We have used an EFT setup which captures all essential and universal features of a light spin-2 field that lives during inflation. Performing SVT decomposition of the spin-2 field, we have found contributions of scalar, vector, and tensor modes to the relic density of the DM. Imposing constraints implied by the backreaction, isocurvature perturbations, BBN and stability, we have shown that the whole DM can be constructed from the spin-2 field. The constraints from the isocurvature perturbations can be easily satisfied when the speed of tensor modes of the spin-2 field is much smaller than the speed of light $\ct\ll1$ while it needs more careful consideration when $\ct={\cal O}(1)$. Finally, assuming instantaneous reheating, we have shown that the whole DM can be produced by the spin-2 particles for the mass range $10^{-3}\, \mbox{eV} \lesssim {m} \ll 10^{5}\, \mbox{GeV}$. The mass range $10^{-3}\, \mbox{eV} \lesssim {m} \ll 10\, \mbox{keV}$ can be only achieved for $c_I\ll1$. 

\vspace{0.7cm}

{\bf Acknowledgments:} I would like to thank Hassan Firouzjahi for critical comments and useful suggestions. I am grateful to Jaume Garriga and Misao Sasaki for stimulating discussions. I also appreciate the people at ICCUB for providing a great environment for free thinking, discussions, and enjoying physics. This work is supported by Mar\'{i}a Zambrano fellowship. 
\vspace{0.7cm}


\bibliographystyle{JHEPmod}
\bibliography{refs}

\end{document}